\documentclass[aps, prb, twocolumn, showpacs, superscriptaddress]{revtex4-1}
\usepackage{lmodern}
\usepackage{amsmath}
\usepackage{physics}
\usepackage{graphicx}
\usepackage{color}
\usepackage{natbib}

\newcommand{\ham}{\ensuremath \mathrm{H}}
\newcommand{\hammat}{\ensuremath \mathbf{H}}
\newcommand{\hamop}{\ensuremath \hat{\ham}}
\newcommand{\cdag}{\ensuremath \hat{c}^\dagger}
\newcommand{\cc}{\ensuremath \hat{c}}
\newcommand{\psibar}{\ensuremath \bar{\psi}_{\alpha E}}
\newcommand{\psitd}{\ensuremath \psi_{\alpha E}}
\newcommand{\psiscat}{\ensuremath \psi^{st}_{\alpha E}}

\begin{document}

\title{Towards Realistic Time-Resolved Simulations of Quantum Devices}

\author{Joseph Weston}
\affiliation{Univ. Grenoble Alpes, INAC-SPSMS, F-38000 Grenoble, France}
\affiliation{CEA, INAC-SPSMS, F-38000 Grenoble, France}
\author{Xavier Waintal}
\affiliation{Univ. Grenoble Alpes, INAC-SPSMS, F-38000 Grenoble, France}
\affiliation{CEA, INAC-SPSMS, F-38000 Grenoble, France}
\date{\today}


\begin{abstract}
We report on our recent efforts to perform realistic simulations of large
quantum devices in the time domain. In contrast to d.c.\ transport where the
calculations are explicitly performed at the Fermi level, the presence of
time-dependent terms in the Hamiltonian makes the system inelastic so that it
is necessary to explicitly enforce the Pauli principle in the simulations. We
illustrate our approach with calculations for a flying qubit interferometer, a
nanoelectronic device that is currently under experimental investigation.  Our
calculations illustrate the fact that many degrees of freedom (16,700 tight-binding sites in the
scattering region) and long simulation times (80,000 times the inverse
bandwidth of the tight-binding model) can be easily achieved on a local computer.
\end{abstract}

\maketitle

\section{Introduction}
\label{sec:intro}
With the technical progress of low temperature (mK) and high frequency (GHz to
THz) experimental setups, high frequency quantum transport experiments have
recently moved from theory to the lab. In particular, coherent single electron
sources with well defined release times~\cite{dubois_minimal-excitation_2013} or
energies~\cite{feve_-demand_2007} have been demonstrated; a key milestone for
engineering non-trivial propagating quantum states. In parallel, quantum
propagation itself has been observed along quantum Hall edge
states~\cite{ashoori_edge_1992,kamata_voltage-controlled_2010,kumada_edge_2011}
and at terahertz frequencies in carbon nanotubes~\cite{zhong_terahertz_2008}.
Being able to perform computer simulations of these types of experiment is an
important step needed for the development of the field.

Simulations of quantum transport in d.c.\ are now routinely performed. When
going from d.c.\ to the time domain, an important difficulty emerges even in the
non-interacting limit: the necessity to enforce the Pauli principle. Indeed, in
presence of time-dependent terms in the Hamiltonian, the energy of an electron
can change, possibly towards a state that is \emph{already occupied}.
As such processes are strictly forbidden, the many-body character of quantum
transport, which can often be forgotten in d.c.\ transport, must be taken care of
properly.

A standard route for doing so is the non-equilibrium Green's functions (NEGF)
formalism, which has been around since the 90s~\cite{wingreen_time-dependent_1993,jauho_time-dependent_1994}. A direct
numerical integration of the NEGF equations is possible but rather cumbersome
even for non-interacting systems (the focus of this paper). Another approach is
the generalization of the Landauer-B\"uttiker scattering matrix formalism to the
time domain~\cite{Moskalets_scattering_2012}. A last approach is the partition-free
approach~\cite{cini_time-dependent_1980,kurth_time-dependent_2005,stefanucci_nonequilibrium_2013}.
These three approaches are strictly equivalent -- at the mathematical level --
for non interacting (or mean field) models~\cite{gaury_numerical_2014}.

In our opinion, the most transparent and efficient way of dealing with the Pauli
principle is to use the fact that the antisymmetric character of a many-body
wave function is preserved by the quantum dynamics.  To be specific, let us
consider a \emph{finite} system $S$ initially at zero temperature, described by
the following Hamiltonian,
\begin{equation}
\label{eq:hamiltonian_finite}
\hamop(t) =  \sum_{i,j \in S} \ham_{ij}(t) \, \cdag_i \cc_j
\end{equation} 
where $\cdag_i$ ($\cc_j$) is a creation
(destruction) operator for a one-particle state on site $i$ ($j$), and $\ham_{ij}(t)$
are matrix elements of the Hamiltonian, which we will collectively refer to
as the matrix $\hammat(t)$.
The ``sites'' $i$ may label spatial degrees of freedom as well as any internal degrees of freedom such as spin or orbitals.
We suppose that
the time-dependent perturbation starts at $t>0$, so that for $t\le0$ the system
can be characterized by its eigenstates $\phi_\alpha$
\begin{equation}
    \hammat(t=0) \phi_\alpha = E_\alpha\phi_\alpha.
\end{equation}
Introducing the operators $\hat{d}_\alpha = \sum_i [\phi_\alpha]_i\, \cc_i$,
($[\phi_\alpha]_i$ is the component of $\phi_\alpha$ on site $i$)
the many-body state at $t=0$ is simply a slater determinant of all the filled
states at energies lower than the Fermi energy $E_F$:
\begin{equation}
    \ket{\Psi(t=0)} = \prod_{E_\alpha < E_F}  \hat{d}_\alpha^\dagger \ket{0},
\end{equation}
where daggers denote Hermitian conjugation and $\ket{0}$ is the vacuum state.
In the same way, the solution at finite time can be written as
\begin{equation}
    \ket{\Psi(t)} = \prod_{E_\alpha < E_F}  \hat{d}_\alpha^\dagger(t) \ket{0}
\end{equation}
with
\begin{equation}
    \hat{d}_\alpha(t) = \sum_i [\psi_\alpha(t)]_i\, \cc_i,
\end{equation}
where the one-body state $\psi_\alpha(t)$ satisfies the Shr\"odinger
equation,
\begin{equation}
    i\hbar \pdv{t}\psi_\alpha(t) = \hammat(t)\psi_\alpha(t)
\end{equation}
and the initial condition
\begin{equation}
    \psi_\alpha (t=0) = \phi_\alpha.
\end{equation}
In other words, one only needs to evolve all the filled \emph{one-body} states in time
and use these to calculate observables. For instance the average number of particles
on site $i$ is simply given by,
\begin{equation}
    \langle \cdag_i(t) \cc_i(t)\rangle  = \sum_{E_\alpha < E_F}  |\psi_\alpha(t)|_i^2
\end{equation}
The approach developed below (which can be found in
Ref.~\cite{gaury_numerical_2014} and Ref.~\cite{weston_linear-scaling_2015})
follows essentially the above line of thought with 2 caveats: (i) one needs to
extend the reasoning to infinite systems, i.e. systems connected to macroscopic
electrodes. This implies that the filled states now form a continuum. (ii) The
system at $t=0$ can be in an out-of-equilibrium state characterized by the
different electrodes having different (electro-)chemical potentials and
possibly different temperatures. Despite these difficulties we recently
developed the ``source-sink'' algorithm~\cite{weston_linear-scaling_2015}
that scales linearly with the number of degrees of freedom in the scattering
region \emph{and} required simulation time.
This algorithm has been applied recently to various cases including
electronic interferometers~\cite{gaury_dynamical_2014,gaury_.c._2015}, quantum
Hall effect~\cite{gaury_stopping_2014}, normal-superconducting
junctions~\cite{weston_manipulating_2015}, Floquet topological
insulators~\cite{fruchart_probing_2016}, Josephson
junction~\cite{weston_linear-scaling_2015} and the calculation of the quantum
noise of voltage pulses~\cite{gaury_computational_2016}. The favorable scaling
properties are put to good use in the present article, where we treat systems
with up to $16,700$ degrees of freedom in the scattering region simulated up
to times of $80,000$ times the inverse bandwidth (the smallest time scale of
the problem). The simulations take just over two hours using a few hundred CPU
cores.

In the rest of this article, we will first review our numerical method with a
fresh emphasis on the summation over filled states. In the final part, we will
simulate an interesting ``flying qubit'' device which is currently the focus of
an important experimental
effort~\cite{bautze_theoretical_2014,takada_transmission_2014,takada_measurement_2015}.

\section{Numerical Method}
\label{sec:numerical}
In this section we will succinctly describe the recently developed source-sink
algorithm mentioned in the previous section. For brevity we shall leave a full
derivation to previously published
works~\cite{gaury_numerical_2014,weston_linear-scaling_2015}.

Let us now consider -- in contrast to section~\ref{sec:intro} -- an \emph{open} quantum system without interactions consisting of a
central scattering region $S$ connected to semi-infinite periodic \emph{leads} $L$ such
that the Hamiltonian can be written as
\begin{equation}\label{eq:hamiltonian}\begin{split}
    \hamop(t) =  \sum_{i,j \in S} &\ham_{ij}(t) \, \cdag_i \cc_j
    \;+ \sum_{i,j \in L} \ham_{ij}(t) \, \cdag_i \cc_j \;+\\
    \sum_{i\in S, j \in L} &\ham_{ij} \, \cdag_i \cc_j + h.c.
\end{split}\end{equation}
where $\ham_{ij}(t)$ are now elements of an \emph{infinite} matrix~$\hammat(t)$.
Note that even if
the leads contain some uniform time-dependent voltage, the Hamiltonian can
always be brought into the form of eq.~\eqref{eq:hamiltonian} by an appropriate
gauge transformation. As in section~\ref{sec:intro} we shall make the restriction that the time-dependent perturbations start at $t>0$
, so that for $t\le0$ the system can be
characterized by its \emph{scattering states} $\psiscat$ labelled by their energy E
and incoming channel $\alpha$ in the leads:
\begin{equation}
    \label{eq:tise}
    \mathbf{H}(t=0)\psiscat = E\psiscat.
\end{equation}
Using the periodic structure of the scattering states in the leads transforms
eq.~\eqref{eq:tise} into a linear system that can be solved using efficient
techniques~\cite{wimmer_quantum_2009,rungger_algorithm_2008}; we use the
Kwant~\cite{groth_kwant:_2014} quantum transport package to obtain the
$\psi^{st}_{E\alpha}$. Once we have the scattering states we can define the
time-evolved scattering states using the time-dependent Schr\"o\-dinger equation:
\begin{equation}
    \label{eq:tdse}
    i \pdv{t} \psitd(t) = \mathbf{H}(t)\psitd(t)
\end{equation}
with the initial condition
\begin{equation}
    \psitd(0) = \psiscat.
\end{equation}
As eq.~\eqref{eq:tdse} is defined on the full, \emph{infinite} domain, it is not
very useful for direct numerical simulation. In  the source-sink
algorithm, one replaces eq.~\eqref{eq:tdse} with a different - yet equivalent - problem in order
to obtain the time-evolved scattering states in the central region. This
consists of solving the following differential equation:
\begin{equation}
    \label{eq:source-sink}
    i \pdv{t} \psibar(t) =
    \left[\mathbf{H}(t) - E\right]\psibar(t) + S(t) + \mathbf{\Sigma}\psibar(t)
\end{equation}
where $\psitd(t) = [\psibar(t) - \psiscat]e^{-iEt}$, $S(t) =
\mathbf{H}(t)\psiscat$ is the source term, and $\mathbf{\Sigma}$ -- the sink -- is a diagonal
matrix that is $0$ in the central region and takes complex values in a finite
portion of the leads. Equation~\eqref{eq:source-sink} is solved over the scattering region plus this
finite portion of the leads, with the initial condition $\psibar(0) = 0$. Once
we have the time-evolved scattering states we can calculate the thermal
averages of physical quantities by integrating over the scattering states that
were occupied at $t=0$. The current between sites $i$ and $j$, for example, can
be written as
\begin{equation}
    \label{eq:current}
    I_{ij}(t) = \sum_\alpha \int \frac{\dd{E}}{2\pi}\, f_\alpha(E) I_{\alpha; ij}(E, t),
\end{equation}
where
\begin{equation}
    \label{eq:energy-resolved-current}
    I_{\alpha;ij}(E, t) = 2\Im\qty( [\psitd^\dagger(t)]_i \, \ham_{ij}(t) \, [\psitd(t)]_j),
\end{equation}
and $f_\alpha(E)$ is the Fermi-Dirac distribution for the lead that contains mode $\alpha$. 
More generally the Retarded and Lesser Green's functions ($G_{ij}^R(t, t')$ and $G_{ij}^<(t, t')$)
can be computed using:
\begin{align}\label{eq:greens}
    G_{ij}^<(t, t') &=
        \sum_\alpha \int \frac{\dd{E}}{2\pi}\, f_\alpha(E)
        [\psitd(t)]_i \; [\psitd^\dagger(t')]_j\\
    G_{ij}^R(t, t') &=
        -i\Theta(t - t')\sum_\alpha \int \frac{\dd{E}}{2\pi}\,
        [\psitd(t)]_i \; [\psitd^\dagger(t')]_j
\end{align}
where $\Theta(t - t')$ is the Heaviside function.

\subsection{Noise properties}
We can even go beyond simple one-particle observables and look at two-particle
observables such as current correlations and noise~\cite{gaury_computational_2016}.
This is possible because the Hamiltonian is quadratic and we can
therefore use Wick's theorem to express these observables as products of
the single-particle Green's functions. Here we shall explicitly show the
expressions for the current noise.

We define the current-current correlation function as
\begin{equation}
    S_{\mu \nu}(t,t') = \left\langle\left(\hat{I}_{\mu}(t) - \langle \hat{I}_{\mu}(t) \rangle \right)
              \times \left(\hat{I}_{\nu}(t') - \langle \hat{I}_{\nu}(t') \rangle \right)\right\rangle,
\label{defcor}
\end{equation}
with the operator for current flowing across an interface $\mu$ defined as
\begin{equation}
\hat{I}_{\mu}(t) = \sum_{\langle i,j \rangle \in \mu} 
\ham_{ij}(t) c_i^{\dagger}(t)c_j(t) - \ham_{ji}(t) c_j^{\dagger}(t)c_i(t),
\end{equation}
where the sum is performed over pairs of sites on opposite sides of the
interface. Expressing this quantity in terms of the scattering
states of the sytem, one arrives at
\begin{equation}
\label{correlator}
\begin{split}
    S_{\mu \nu}(t,t') =
    \sum_{\alpha, \beta} \int \frac{dE}{2\pi} \int \frac{dE'}{2\pi}
    &f_{\alpha}(E)(1-f_{\beta}(E')) \,\times\\
    &I_{\mu,EE'}(t) \qty[I_{\nu,EE'}(t')]^*,
\end{split}
\end{equation}
where the quantity $I_{\mu,EE'}(t)$ is closely related to the initial current operator
\begin{equation}
\label{EEcurrent}
\begin{split}
    I_{\mu,EE'}(t) = \sum_{\langle i,j \rangle \in \mu}\Big(
    &[\psi_{\beta E'}^\dagger(t)]_i \, \ham_{ij}(t) \, [\psi_{\alpha E}(t)]_j \,-\\
    &[\psi_{\beta E'}^\dagger(t)]_j \, \ham_{ji}(t) \, [\psi_{\alpha E}(t)]_i \Big).
\end{split}
\end{equation}
Equation~\eqref{correlator} relates the typical output of a time-dependent
simulation (right-hand side) to the noise  properties (left-hand side). As an
example, the total number of transmitted particles $\hat n_\mu$ over a duration
$\Delta$, defined as

\begin{equation}
    \hat{n}_\mu = \int_{-\Delta/2}^{\Delta/2} dt\ \hat{I}_{\mu}(t),
\end{equation}
can be calculated from the above expression and one arrives at a simple closed expression in 
terms of the time-dependent wave functions:
\begin{equation}
\label{var3}
    \mathrm{var}(\hat{n}_{\mu}) =  \sigma^2_{st} \ \Delta  +  2\sigma_{mix} + \bar \sigma^2  + O(1/\Delta)
\end{equation}
with
\begin{align}
    \sigma^2_{st} &= \sum_{\alpha, \beta} \int \frac{dE}{2\pi}\ f_{\alpha}(E)(1-f_{\beta}(E))
    |I_{\mu,EE}(0)|^2 \\
    \sigma_{mix} &= \sum_{\alpha, \beta} \int \frac{dE}{2\pi}\ f_{\alpha}(E)(1-f_{\beta}(E))
     \mathrm{Re}[\bar{N}_{EE}^* I_{\mu,EE}(0)] \\
    \bar \sigma^2 &= \sum_{\alpha, \beta} \int \frac{dE}{2\pi}\ \frac{dE'}{2\pi}
    f_{\alpha}(E)[1-f_{\beta}(E')] |\bar{N}_{EE'}|^2
\end{align}
with $\bar{N}_{EE'}$ defined as
\begin{align}
    \bar{N}_{EE'} = \int_{-\infty}^{\infty} dt\ \left[I_{\mu,EE'}(t) -
    I_{\mu,EE'}(0)e^{-i(E-E')t}\right].
\end{align}

\subsection{Performing the Energy Integral}
\label{sec:energy-integral}
The above is a complete prescription for calculating the expectation value of
time-resolved quantities in non-interacting nanoelectronics systems with semi-infinite leads.  
While eq.~\eqref{eq:source-sink} can be solved numerically
using standard techniques (we use a Runge-Kutta-Fehlberg adaptive
scheme~\cite{fehlberg_Klassische_1970}), the energy integration needed to
compute the thermal average can require a little more care. In particular, the
integrands of expressions such as eq.~\eqref{eq:current} typically have
(integrable) divergences at energies where the bands open.  This is because the
scattering states are normalized so that they carry unit current (so that
the scattering matrix is unitary), which means that they diverge as
$1/\sqrt{v_\alpha(E)}$, where $v_\alpha(E) = \dd{E_\alpha}/\dd{k}$ is the group velocity of mode
$\alpha$ at energy $E$. For one-particle observables such as the current this
means that the integrand diverges as $1/v_\alpha(E)$ near the energy $E_\alpha$
at which the mode $\alpha$ opens (although as $v_\alpha(E) \propto \sqrt{E}$
near the band edge, this divergence is integrable). This can be solved simply
by choosing to integrate in momentum instead of energy; the $\dd{E}/\dd{k}$
Jacobian factor cancels the divergence. When integrating in momentum one
must make sure to only integrate over regions where $v_\alpha(k) > 0$, as these
states are \emph{incoming} into the system (i.e. they correspond to our
scattering states). Equations~\eqref{eq:current}
and~\eqref{eq:energy-resolved-current} can thus be rewritten as
\begin{equation}
    I_{ij}(t) = \sum_\alpha \int_{-\pi}^{\pi} \frac{\dd{k}}{2\pi}\, f_\alpha(k)\,
    \Theta[v_{\alpha}(k)]\, v_{\alpha}(k)\, I_\alpha(k, t),
\end{equation}
and
\begin{equation}
    I_\alpha(k, t) = 2\Im\qty([\psi_{\alpha k}^\dagger(t)]_i \, \ham_{ij}(t) \, 
    [\psi_{\alpha k}(t)]_j).
\end{equation}
\begin{figure}
    \centering
    \includegraphics[width=0.5\textwidth]{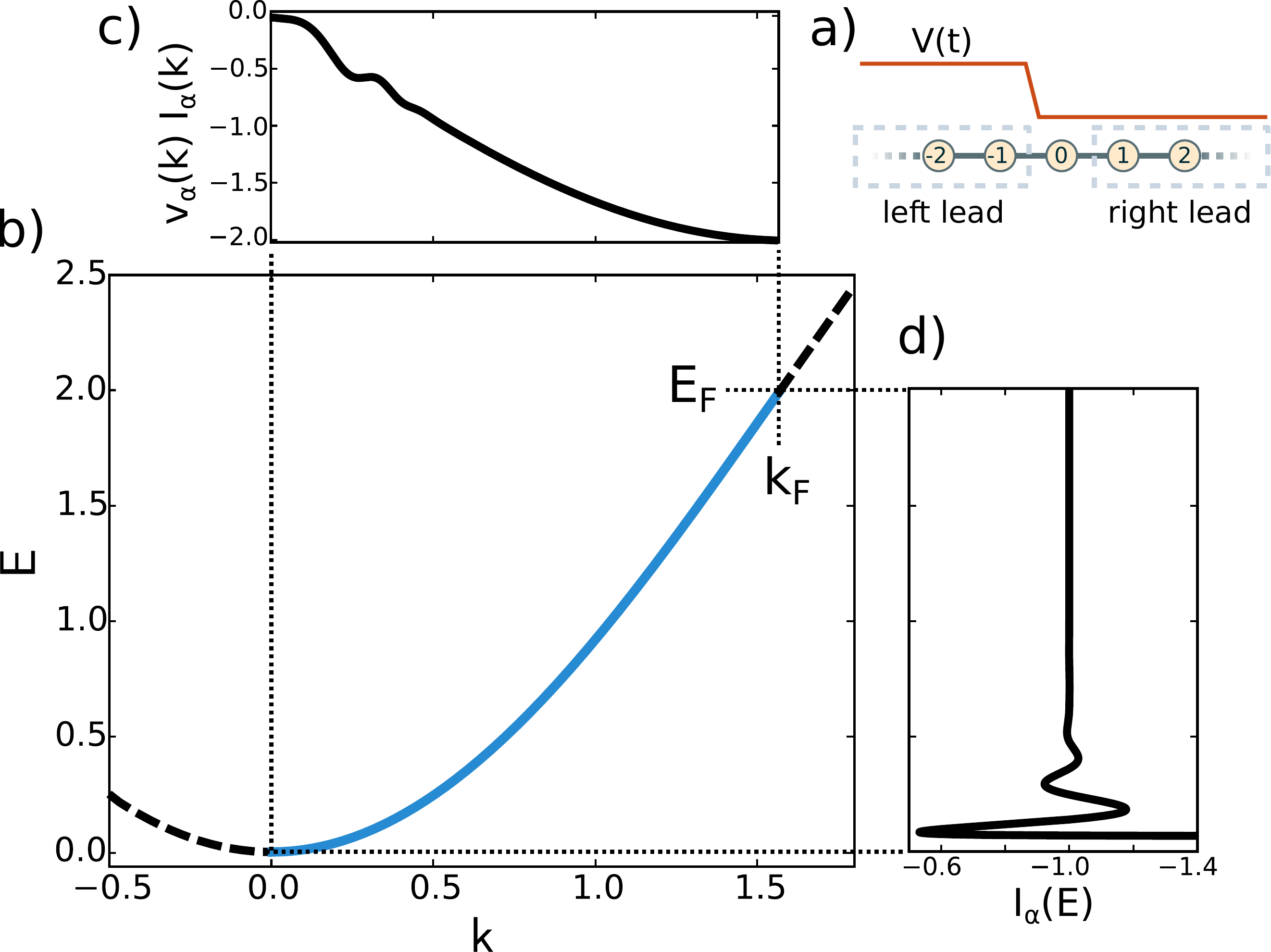}
    \caption{An illustration of the difference between the energy and momentum
        integration techniques applied to a 1D chain with a gaussian voltage pulse
        applied. Times are given in units of the inverse bandwidth.
        a) Sketch of the simulated system. b) The band structure for the left lead.
        c) Integrand in k-space to calculate the current at $t=100$.
        d) Integrand in energy to calculate the current at $t=100$.
    }
    \label{fig:integration_scheme}
\end{figure}
Figure~\ref{fig:integration_scheme} shows the calculated $I_\alpha(E, t)$ and
$v_\alpha(k) I(k, t)$ at $t = 100$ (in units of the inverse bandwidth) for the mode coming from the
left-hand lead in a perfect 1D chain after the application of a voltage pulse
on the left-hand lead. We clearly see the $1/v_{\alpha}(E)$ divergence in $I_\alpha(E)$
at $E=0$ in fig.~\ref{fig:integration_scheme}d, which is regularized by the
change of variables to $k$, shown in fig.~\ref{fig:integration_scheme}c.

\begin{figure}
    \centering
    \includegraphics[width=0.5\textwidth]{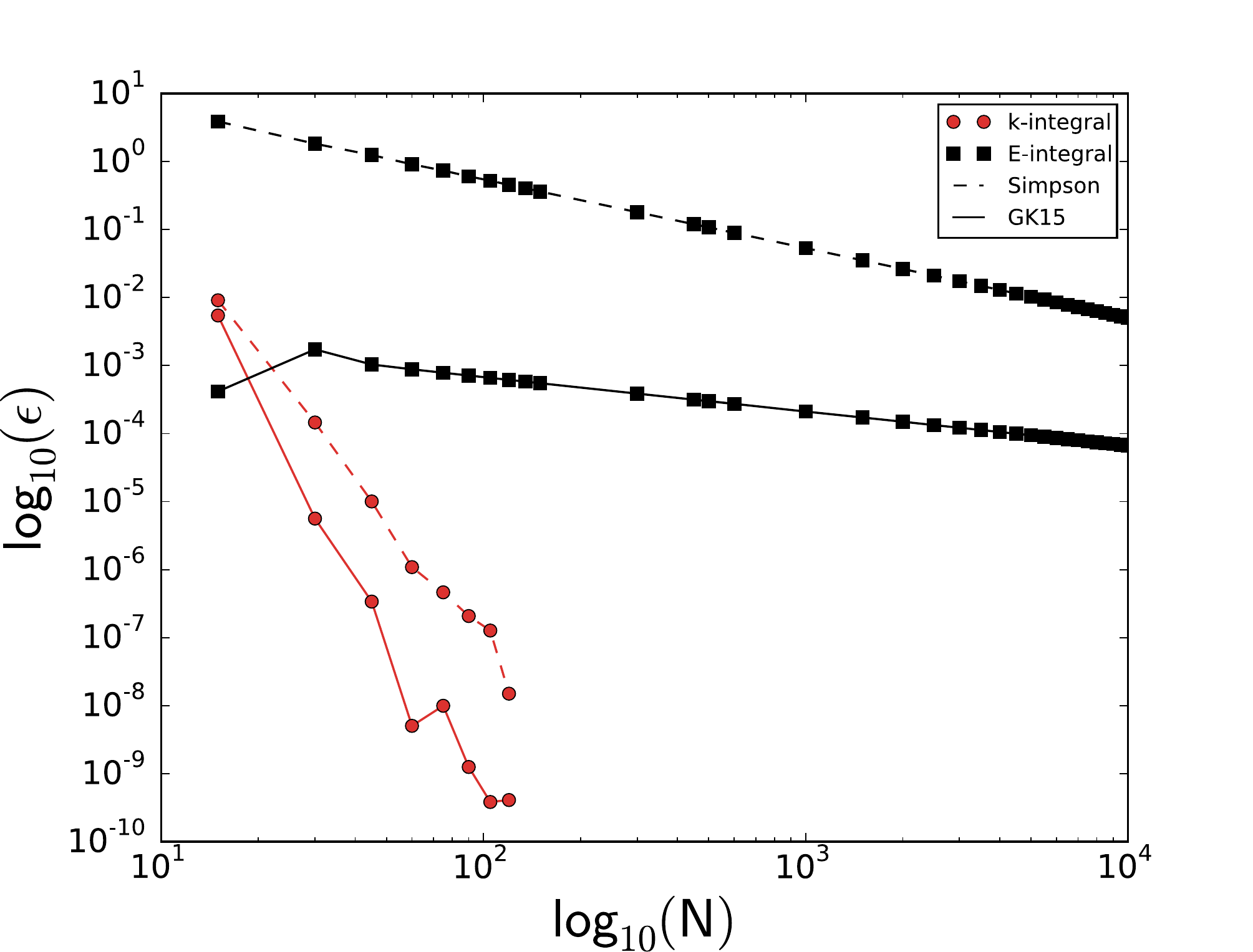}
    \caption{Comparison of the error $\epsilon$ for integration along $k$ and
        energy using different integration methods -- Simpson's rule and
        Gauss-Kronrod 15(7) (GK15) -- as a function of the number of integrand
        evaluations, $N$. The error was calculated by subtracting the results
        of the integrals from a reference calculation performed using a k-space
        integration and the GK15 rule with 667 subregions (10005 integrand
        evaluations).
    }
    \label{fig:integration_error}
\end{figure}

Once the integration regions have been defined the integral is computed by
using a Gauss-Kronrod 15(7) point embedded
scheme~\cite{piessens_quadpack_1983}. Figure~\ref{fig:integration_error}
compares the error scaling as a function of the number of integrand evaluations
for this Gauss-Kronrod rule and Simpson's rule, applied to both the energy and
k-space integrations, of the integrands shown in
fig.~\ref{fig:integration_scheme}.  We clearly see that the k-space integral
has significantly better scaling in this case due to the regularized
singularity at the band edge. In addition, we see that the Gauss-Kronrod rule
is capable of achieving orders of magnitude greater accuracy for a given number
of integrand evaluations. We note that we can benefit from this reduction
only because our problem is formulated in the continuum, and so we are calculating
\emph{integrals} (as opposed to discrete sums) that can be evaluated numerically
by discretizing in any way we please.  If instead we had started from a
(possibly large) finite system, the choice of $k$ points would have been
dictated by the manner in which we truncated the infinite system.

An additional advantage of embedded rules such as Gauss-Kronrod is that they
allow an error estimate to be calculated with no extra integrand evaluations.
The integration region can then be bisected and the integral re-computed on the
subregions if the error is found to be unacceptably high.  Also, such schemes
have the advantage that one does not need to evaluate the integrand on the
boundary, as one does for schemes such as Simpson's rule. This is advantageous
for our purposes as the scattering problem is ill-conditioned at energies where
new bands open, so one needs to add an artificial offset to the integration
boundaries when using a scheme that requires integrand evaluations
on the boundary.  The problem of using an adaptive scheme is that re-computing
the integrand is computationally expensive (it corresponds to re-evolving
wavefunctions from $t=0$ up to the time at which we wish to calculate the
observable). For the specific case of systems that only have time-dependence in
the voltage in the leads, we have found that a useful technique is to already
subdivide the integration regions at $t=0$, so as to obtain a good estimate for
the observable at that time (when computing the integrand is computationally
cheap; there is no time evolution to do!).  At later times the integral is
typically well-estimated by using this initially-chosen set of subintervals,
and requires fewer additional subdivisions. In addition, any known structure of
the problem (e.g. positions of resonances etc.) can also aid in an effective
initial choice of subregions.

\section{Application to a Flying Qubit Interferometer}
We shall now apply the source-sink algorithm to a flying qubit interferometer
in a split-wire setup. Such a setup has recently been realized
experimentally~\cite{bautze_theoretical_2014} and was also studied numerically
in d.c., however here we shall perform time-resolved simulations of a
charge pulse injected into the interferometer by a voltage pulse applied to
one of the contacts.

\begin{figure}
    \centering
    \includegraphics[width=0.5\textwidth]{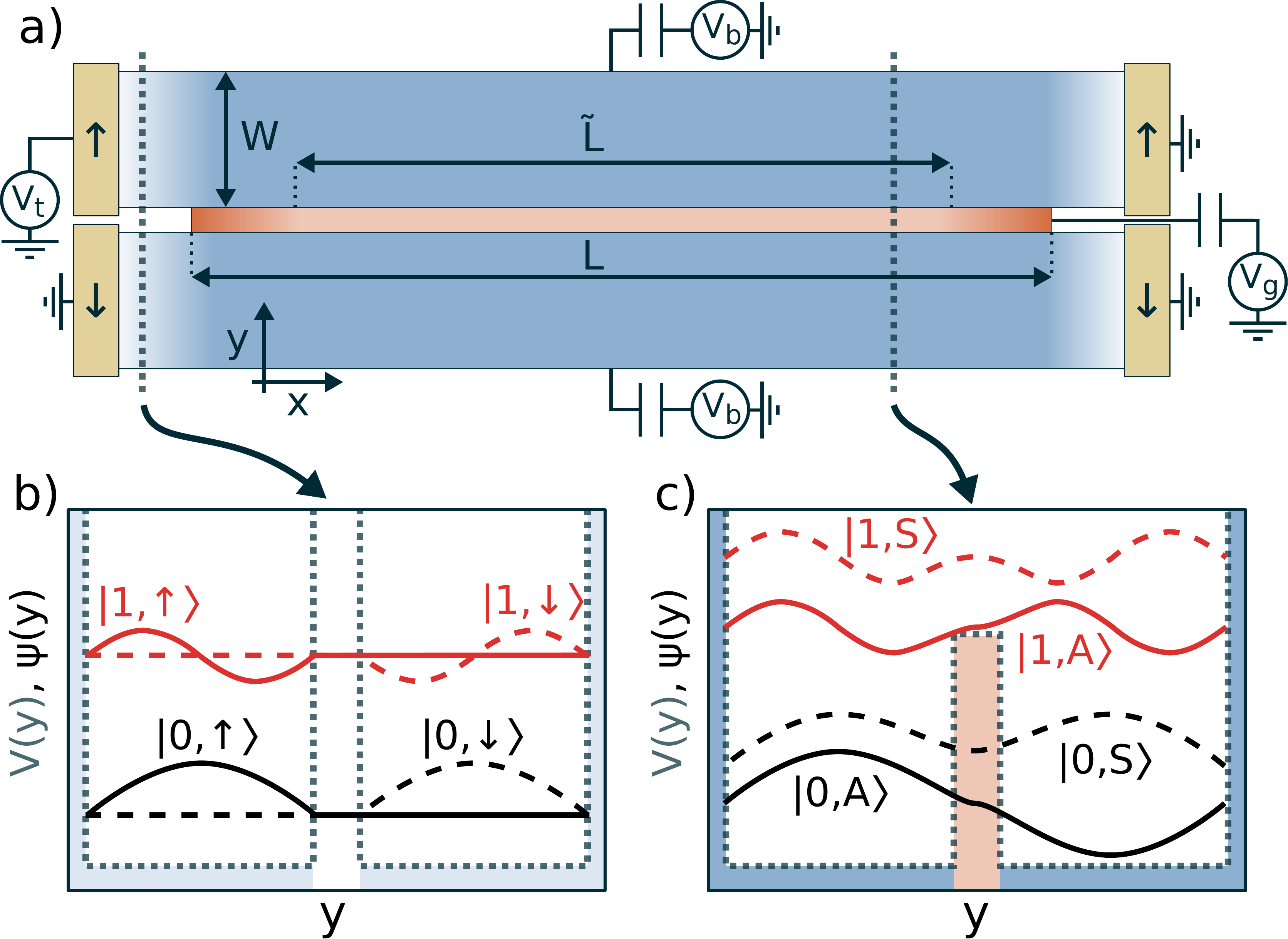}
    \caption{
        a) Sketch of the split-wire system with the coupling region
        in red (controlled by gate voltage $V_g$), a global backgate controlled
        by $V_b$, and 4 leads  which are all grounded except
        for left lead $\uparrow$, which has a (time-dependent) bias voltage $V_t$ applied.
        $L$ measures the total coupling region length, $\tilde{L}$ measures the effective
        coupling region length, and $W$ measures the width
        of an individual wire.
        b) Sketch of the 4 lowest energy transverse modes before the coupled region.
        along with the transverse potential, $V(y)$. The states $\ket{n,\uparrow}$
        and $\ket{n,\downarrow}$ are degenerate.
        c) Sketch of the 4 lowest energy transverse modes in the coupled region,
           along with the transverse potential, $V(y)$.
    }
    \label{fig:split-wire}
\end{figure}

A sketch of the setup is shown in fig.~\ref{fig:split-wire}.  We shall treat
the system as two quasi-1D wires that lie parallel to one another. The two
wires are labelled $\uparrow$ and $\downarrow$ and can be interpreted as the two states
of a (flying) qubit. They are
(rather weakly) connected only in a finite region of length $L$ (shown as the red region in fig.~\ref{fig:split-wire}). A negatively
polarized top-gate placed over the center of the wire in the coupling region
allows the coupling between the wires to be tuned by altering the gate voltage,
$V_g$. Under the gate ($x\approx 0$), the potential falls from $\infty$ to
$V_g$ adiabatically (to avoid spurious reflection), so that the effective
length of the coupling region is $\tilde{L}$.  In addition there is a back-gate
at voltage $V_b$ placed over the whole coupling region that allows us to control the
potential there and hence the number of open conducting
channels. This potential falls adiabatically to 0 before reaching the leads
(shown as the blue color gradient in fig.~\ref{fig:split-wire}).  There is
also a voltage source attached to lead $\uparrow$ on the left that can apply a time-dependent bias
$V_t$ to the system.  We treat the voltage drop as being abrupt at the
interface between the lead $\uparrow$ on the left and the central region. In
order to simulate this model using the source-sink algorithm we first
discretize it onto a square lattice of spacing $a$ in order to obtain a
tight-binding Hamiltonian of the form eq.~\eqref{eq:hamiltonian}.

\subsection{d.c.\ Characterization}
We shall now look at d.c.\ transport in this system, in order to show that it
can be thought of as an interferometer. This point of view will be invaluable
when interpreting the time-resolved simulations in
section~\ref{sec:time-resolved}. In order to calculate the d.c.\ conductance
$G_{\sigma'\,\sigma}$ between lead $\sigma$ on the left and lead $\sigma'$ on the right
($\sigma\in\{\uparrow,\downarrow\}$) we need only use the Landauer
formula~\cite{landauer_spatial_1957,landauer_electrical_1970}:
\begin{equation}
    \label{eq:landauer}
    G_{\sigma'\,\sigma} = \frac{2e^2}{h} D_{\sigma'\,\sigma}
\end{equation}
where $D_{\sigma'\,\sigma}$ is the transmission from lead $\sigma$ on the left
to lead $\sigma'$ on the right, defined by
\begin{equation}
    D_{\sigma'\,\sigma} = \sum_{n,m} T_{m\sigma',n\sigma}
\end{equation}
where $T_{m\sigma',n\sigma}$ is the transmission \emph{probability} from
mode $\ket{n,\sigma}$ on the left to mode $\ket{m,\sigma'}$ on the right
(these modes are sketched in fig.~\ref{fig:split-wire}). In all that
follows we shall assume that inter-band scattering is negligible, i.e.
$T_{m\sigma',n\sigma} = \delta_{mn} T_{n\sigma',n\sigma}$ where $\delta_{mn}$
is the Kronecker delta.
The calculation of the
transmission probabilities for this system was very well explained in section III.B of
ref.~[\onlinecite{bautze_theoretical_2014}], so here we will provide just an intuitive
picture of what is happening.
The full wavefunction in the uncoupled region can be written $\Psi_{n,
\sigma}(x, y) = \braket{y}{n,\sigma}e^{ik_{n,\sigma}x}$, where $\sigma \in
\{\uparrow, \downarrow\}$, and its energy is $E = E_{n,\sigma} +
(\hbar^2/2m^*)k_{n,\sigma}^2$, where $E_{n,\sigma}$ is the energy of the
transverse mode $\ket{n, \sigma}$.  As the states $\ket{n, \uparrow}$ and
$\ket{n, \downarrow}$ are degenerate for a given $n$, we can also define
symmetric and antisymmetric superpositions:
\begin{equation}\begin{split}
    \ket{n, \uparrow} &= (1/\sqrt{2})\qty[\ket{n, S_u} + \ket{n, A_u}]\\
    \ket{n, \downarrow} &= (1/\sqrt{2})\qty[\ket{n, S_u} - \ket{n, A_u}],
\end{split}\end{equation}
where the $u$ subscript reminds us that these are transverse modes in the
\emph{uncoupled} region. In the coupled region we also have symmetric and
antisymmetric modes $\ket{n, S}$ and $\ket{n, A}$, and we suppose that the
transition from the uncoupled to the coupled region is adiabatic, such that
$\ket{n, S_u}$ evolves into $\ket{n, S}$ and $\ket{n, A_u}$ evolves into
$\ket{n, A}$ with no inter-mode scattering. While $\ket{n, A_u}$ and $\ket{n,
S_u}$ are degenerate, $\ket{n, A}$ and $\ket{n, S}$ are not. This means that
for a given energy the states will have different longitudinal wavevectors,
$k_{n, A}$ and $k_{n, S}$. If we were initially in a state $\ket{n, \uparrow}$
this means that a length $\tilde{L}$ after the wires are coupled we will be
in a state:
\begin{equation}
    \ket{\psi_{n, \uparrow}} = \frac{1}{\sqrt{2}}
    \qty[e^{ik_{n, A}\tilde{L}}\ket{n, A} + e^{ik_{n, S}\tilde{L}}\ket{n, S}].
\end{equation}
The wires are then adiabatically uncoupled (near the
right-hand leads) and we can write:
\begin{equation}\begin{split}
    \ket{\psi_{n, \uparrow}} = \frac{1}{2} \bigl[
        &\qty(e^{ik_{n, S}\tilde{L}} + e^{ik_{n, A}\tilde{L}})\ket{n, \uparrow} +\\
        &\qty(e^{ik_{n, S}\tilde{L}} - e^{ik_{n, A}\tilde{L}})\ket{n, \downarrow}
    \bigr].
\end{split}\end{equation}
We immediately see that the difference in wavevectors will give rise to
interference between the symmetric and antisymmetric components.  We can thus
write down the transmission \emph{amplitudes} for arriving on the right in
$\ket{n, \uparrow}$ or $\ket{n, \downarrow}$ given that we were
injected on the left in $\ket{n, \uparrow}$:
\begin{equation}\label{eq:transmission}\begin{split}
    t_{n\uparrow,n\uparrow} &=
        \exp\qty(i\frac{k_{n, A} + k_{n, S}}{2}\tilde{L}) \cos\qty(\frac{\Delta k_n}{2}\tilde{L})\\
    t_{n\downarrow,n\uparrow} &=
        i\exp\qty(i\frac{k_{n, A} + k_{n, S}}{2}\tilde{L}) \sin\qty(\frac{\Delta k_n}{2}\tilde{L}),
\end{split}\end{equation}
where $\Delta k_n = k_{n, A} - k_{n, S}$.  The transmission probabilites can be
calculated from these amplitudes using
$T_{n\sigma',n\sigma} = \abs{t_{n\sigma',n\sigma}}^2$.

\begin{figure}
    \centering
    \includegraphics[width=0.5\textwidth]{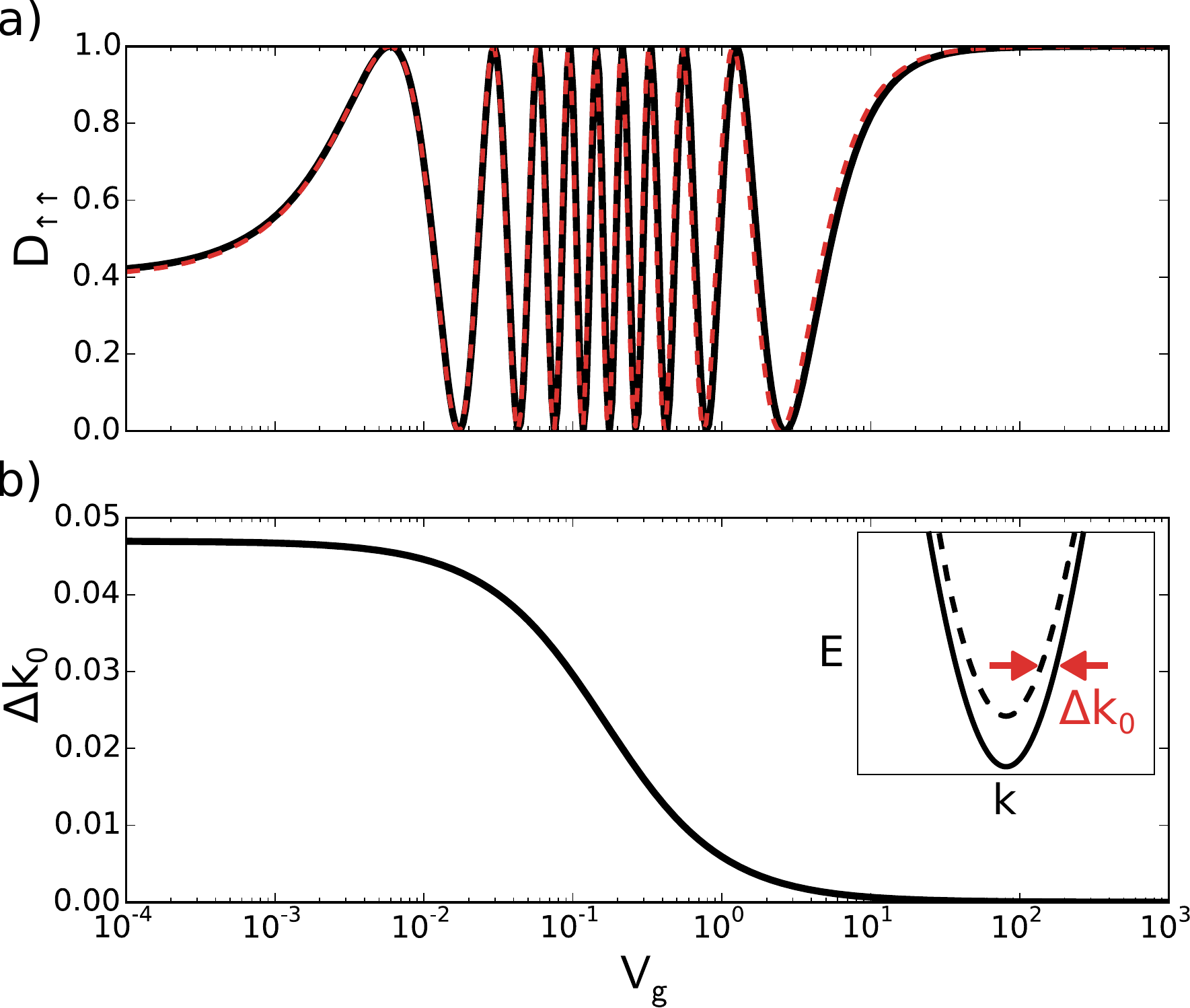}
    \caption{d.c.\ simulation of split wire discretized with lattice parameter
        $a=2$ with $L=1400$, $W=20$, $V_b=0.11$ and $E_F=0.15$. At this Fermi
        energy only the modes $\ket{0, \uparrow}$ and $\ket{0, \downarrow}$ in
        the coupled region are open. a) Black full line: transmission calculated
        from tight-binding simulation, red dashed line: transmission calculated
        using eq.~\eqref{eq:transmission} with $\Delta k_0$ calculated from
        tight-binding and $\tilde{L}$ as a fitting parameter. We used $\tilde{L} = 1242$.
        b) Tight-binding calculation of the difference in momentum between
        symmetric and antisymmetric modes in the coupling region. Both plots
        share the x-axis $V_g$ scale.
    }
    \label{fig:T_Vg}
\end{figure}

We performed tight binding simulations of the d.c.\ split wire system using the
Kwant~\cite{groth_kwant:_2014} package.  Figure~\ref{fig:T_Vg} shows how the
wavevector difference changes as a function of the coupling gate voltage $V_g$ and the effect
that this has on the transmission $D_{\uparrow\,\uparrow}$ from lead $\uparrow$
on the left to lead $\uparrow$ on the right. We
clearly see regular oscillations when the wavevector difference changes
linearly. As we go to to very high gate voltages we effectively uncouple the
two wires, which explains why $D_{\uparrow\,\uparrow} \to 1$ in this limit. The red dashed line
in fig.~\ref{fig:T_Vg}a shows $D_{\uparrow\,\uparrow}$ calculated using
eq.~\eqref{eq:transmission} where $\Delta k_0$ has been calculated
numerically from the tight binding model. We see a good fit of this simplified
model with the full tight binding simulation.

\begin{figure*}[tb]
    \centering
    \includegraphics[width=\textwidth]{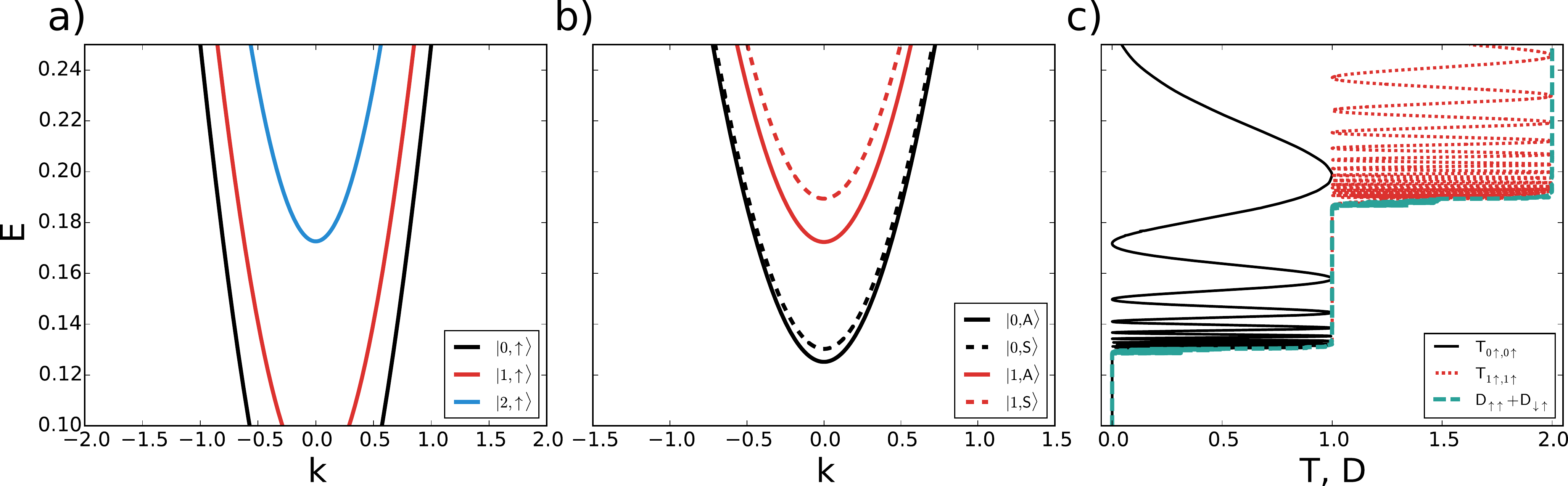}
    \caption{
        Dispersion relations and transmissions for the split wire discretized
        with lattice parameter $a=2$ with $L=1400$, $W=20$, $V_g=0.27$ and
        $V_b=0.11$.
        a) Dispersion relation in lead $\uparrow$ showing the three lowest
        energy modes.
        b) Dispersion relation in the coupling region of the split wire,
        showing the first 2 symmetric (full lines) and anti-symmetric (dashed lines) modes.
        c) Transmission probability from $\ket{0,\uparrow}$ on the left to
        $\ket{0, \uparrow}$ on the right (black full line); transmission probability from
        $\ket{1,\uparrow}$ on the left to $\ket{1, \uparrow}$ on the right
        (red dotted line, shifted by 1 for clarity); total transmission from the $\uparrow$
        lead on the left to the leads on the right (green dashed line).
        All three plots share the y-axis energy scale.
    }
    \label{fig:T_Ef}
\end{figure*}

Figure~\ref{fig:T_Ef} shows the dispersion relations for the leads
(subfigure a) and in the coupled region (subfigure b) calculated from
the tight-binding model. We see in fig.~\ref{fig:T_Ef}c the transmission
probabilities for being transmitted through the first and second modes
from lead $\uparrow$ on the left to lead $\uparrow$ on the right.
We see that the transmission probabilities are 0 before the
corresponding modes in the central region open. Note that
in order for $T_{n\sigma',n\sigma}$ to be different from 0 we need
\emph{both} modes $\ket{n,A}$ \emph{and} $\ket{n,S}$ to be open in the
coupled region, as $\ket{n, \sigma}$ is a linear combination of
both. We see that the transmission probabilities oscillate as a function
of energy. The reason for this is clear, as fig.~\ref{fig:T_Ef}b clearly
shows that $\Delta k_n$ changes as a function of energy.
The inter-band transmission probabilities $T_{m\sigma',n\sigma}$ (with
$m \ne n$) are not shown, but are 0 at all energies (validating the assumptions
of the analytical derivation above); this is because
the transition from uncoupled to coupled region is done in an adiabatic
manner.

One last point, which is perhaps a bit subtle, is that we expect to be able to
see these interference effects even with a \emph{large} number of open
channels. Indeed, at the energy where the $n+1$ channel
opens the $\Delta k_{n+1}$ is much larger than the $\Delta k_n$ at the same
energy (see fig.~\ref{fig:T_Ef}b). This means that $T_{n\sigma',n\sigma}$ oscillates much slower than
$T_{n+1\,\sigma',n+1\,\sigma}$ at the same energy, as can be clearly seen in
fig.~\ref{fig:T_Ef}c. This separation in frequency of the different
$T_{n\sigma',n\sigma}$ means that the oscillations from the different channels
will be clearly distinguishable in the full differential conductance $G_{\sigma'\,\sigma}$.

\subsection{Time-Dependent Simulations}
\label{sec:time-resolved}
Now we shall turn to time-resolved simulations of the split-wire. We apply a
Gaussian voltage pulse to lead $\uparrow$ on the left and measure the current $I_\uparrow$ ($I_\downarrow$) leaving the system on the right via lead $\uparrow$ ($\downarrow$). We
also measure the injected current $I_{in}$ .  We place ourselves in a regime
where only the modes $\ket{0, A}$ and $\ket{0, S}$ are open in the coupled
region at the Fermi energy, and the pulse is not short/intense enough to excite
higher energy modes. In d.c.\ this system has transmission $D_{\uparrow\,\uparrow}=0.1$
and $D_{\downarrow\,\uparrow} = 0.9$. Our system has $L=1400$ and $W=20$ with a
discretisation parameter $a=2$, and in total we have $16,700$ sites in the
scattering region.

\begin{figure*}
    \centering
    \includegraphics[width=\textwidth]{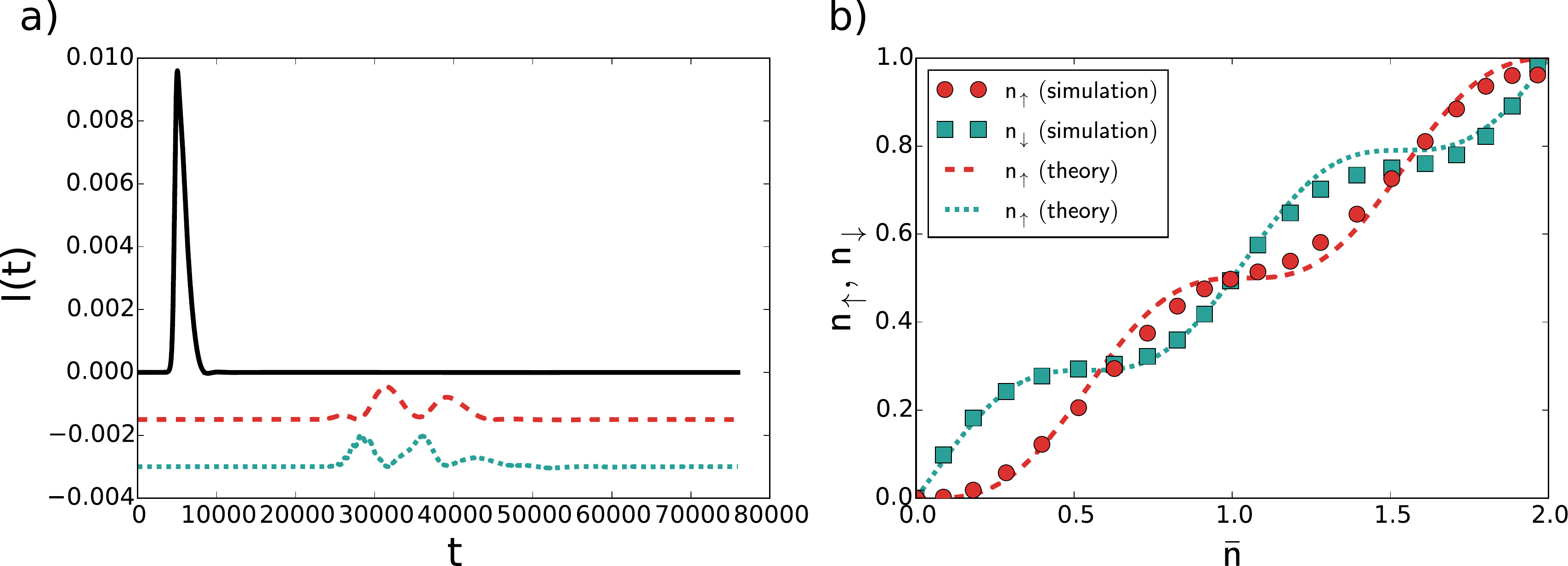}
    \caption{Charge transport after application of a voltage pulse on lead $\uparrow$
        on the left of the split-wire. a) Current as a function of time flowing in:
        lead $\uparrow$ on the left,
        ($I_{in}$), lead $\uparrow$ on the right ($I_\uparrow$), and lead
        $\downarrow$ on the right($I_\downarrow$), for a voltage
        pulse that injects $\bar{n} = 2.0$ particles. b)
        Number of transmitted particles on the right in lead $\uparrow$
        ($n_\uparrow$) and lead $\downarrow$ ($n_\downarrow$) as a function
        of the injected number of particles
        ($\bar{n}$). Symbols: time-resolved simulation, dashed line: application
        of eq.~\eqref{eq:td_transmission} with $\tilde{L}=1340$ as a fitting parameter.
        The system is discretized with lattice parameter $a=2$ with
        $L=1400$, $W=20$, $V_b=0.11$, $V_g=0.1446$ and $E_F=0.15$. We use a pulse
        with a duration (full-width half-maximum) of $1600$ times the inverse bandwidth. This setup
        gives a d.c.\ transmission of 0.9 from lead~$\uparrow$ on the left to
        lead~$\downarrow$ on the right at the Fermi energy.
    }
    \label{fig:nt_nbar}
\end{figure*}

Figure~\ref{fig:nt_nbar}a shows the results of a
simulation where the above-defined currents are measured. Due to the large characteristic length $1/\Delta k_{0}$,
hence the large length of the system, we need to go to very long times ($80,000$ times
the inverse bandwidth) in order to see the output current. The voltage pulse
shown injects an average of $\bar{n}=2$ particles into system, where
\begin{equation}
    e\bar{n} = \int_0^\infty \dd{t}\, I_{in}(t)
\end{equation}
and $e$ is the electronic charge. We clearly see that the output current
oscillates between the $\uparrow$ and $\downarrow$ leads, which is
counterintuitive; na\"ively one would expect that the current in the
two leads would have the same ``shape'' as a function of time, and
that only the magnitudes would be different (proportional to the
d.c.\ transmission).

Figure~\ref{fig:nt_nbar}b shows the the number of
particles transmitted on the right into lead $\uparrow$ ($n_\uparrow$) and
$\downarrow$ ($n_\downarrow$) as
a function of the number of injected particles. Rather than a simple
proportionality relationship (where the slope would be given by the d.c.\
transmission), we see that the number of particles depends non-linearly on  $\bar{n}$
and even \emph{oscillates} with $\bar{n}$. This curious behaviour can be understood by first understanding
that prior to the voltage pulse, the Fermi sea is made of waves that form an effective two-path 
(Mach-Zehnder) interferometer. The effect of the voltage pulse is to twist the phase of these scattering
states, which eventually leads to the  ``dynamical modification of the interference pattern'' previously
calculated in ref.~[\onlinecite{gaury_dynamical_2014}]. The physics of the current
split-wire setup and the Mach-Zehnder interferometer studied in
ref.~[\onlinecite{gaury_dynamical_2014}] is very similar; the only real difference is that
the two paths of the Mach-Zehnder are spatially separated while here they share the same region.  Adapting the
results of ref.~[\onlinecite{gaury_dynamical_2014}] (eqs. 27 and 28 in
ref.~[\onlinecite{gaury_dynamical_2014}]) to the present
case yields
\begin{equation}\label{eq:td_transmission}\begin{split}
    n_\uparrow &= \frac{\bar{n}}{2} + \frac{1}{2\pi}\sin(\pi\bar{n})
        \cos\qty(\pi\bar{n} + \frac{\Delta k_0}{2}\tilde{L})\\
    n_\downarrow &= \frac{\bar{n}}{2} - \frac{1}{2\pi}\sin(\pi\bar{n})
        \cos\qty(\pi\bar{n} + \frac{\Delta k_0}{2}\tilde{L}).
\end{split}\end{equation}
Figure~\ref{fig:nt_nbar}b compares the results of the time-resolved
simulation (symbols) with the above model (dashed lines) with
$\tilde{L}$ as a fitting parameter; we used $\tilde{L}=1340$. We
see a very good agreement of the numerical results with the
theory.

\section{Conclusions}
In the vast majority of devices proposed to implement qubits, one applies time-dependent excitations to a localized two level system in order to create the desired superposition of states. In the device studied above, the philosophy is sligthly different: the time-dependent potential is replaced by a spatially dependent potential and, upon sending a charge excitation inside the system, this excitation travels and \emph{effectively} experiences a time dependent potential as it sees different regions of the sample. In a similar spirit, one could implement two-qubit gates by capacitively coupling two such devices through one arm. Such devices could be used as qubit on their own right, or perhaps more interestingly as quantum buses to couple different localized qubits.

We have performed time-resolved simulations of this split-wire flying qubit setup using cutting-edge numerical techniques. We have seen that such a system can be effectively treated as a two-path interferometer,
and that applying voltage pulses to a lead of the split-wire gives rise to a modification of the d.c.\ interference pattern; a strictly dynamical effect. More physics needs to be included into the model to achieve realistic simulations, starting with the (self consistent) Hartree potential that leads to a renormalization of the surface plasmon velocity. However, the technology is now in place so that these types of simulation will become mature and be used as design tools for future experiments.

\emph{Acknowledgements}.\;
This work was supported by the ANR grant QTERA, and the ERC consolidator grant
MesoQMC.  We thank Christopher Ba\"uerle, Gr\'egoire Roussely and Shintaro
Takada for interesting discussions.

\bibliographystyle{apsrev}       
\bibliography{flying-qubit}

\end{document}